# *No actual measurement … was required:* Maxwell and Cavendish's null method for the inverse square law of electrostatics

Isobel Falconer

School of Mathematics and Statistics
University of St Andrews
North Haugh
St Andrews
Fife
UK
KY16 9SS

Email: ijf3@st-andrews.ac.uk

**Abstract**
In 1877 James Clerk Maxwell and his student Donald MacAlister refined Henry Cavendish's 1773 null experiment demonstrating the absence of electricity inside a charged conductor. This null result was a mathematical prediction of the inverse square law of electrostatics, and both Cavendish and Maxwell took the experiment as verifying the law. However, Maxwell had already expressed absolute conviction in the law, based on results of Michael Faraday's. So, what was the value to him of repeating Cavendish's experiment? After assessing whether the law was as secure as he claimed, this paper explores its central importance to the electrical programme that Maxwell was pursuing. It traces the historical and conceptual re-orderings through which Maxwell established the law by constructing a tradition of null tests and asserting the superior accuracy of the method. Maxwell drew on his developing 'doctrine of method' to identify Cavendish's experiment as a member of a wider class of null methods. By doing so, he appealed to the null practices of telegraph engineers, diverted attention from the flawed logic of the method, and sought to localise issues around the mapping of numbers onto instrumental indications, on the grounds that 'no actual measurement … was required'.

## 1. Introduction
In the first edition of his *Treatise on Electricity and Magnetism* James Clerk Maxwell acknowledged Charles-Augustin Coulomb's 1785 torsion balance experiments as establishing the inverse square law of electrostatics (1873, pp. 41, 43). Within a few pages he dismissed them again as demonstrating the law only to a rough approximation. Instead he cited Faraday's observation that an electrified body, touched to the inside of a conducting vessel, transfers *all* its electricity to the outside surface, as "far more conclusive than any measurements of electrical forces can be" (p. 75). He based this assertion on his mathematical contention that electricity would reside entirely on the outside of a closed conductor – with none inside - *only* if the microscopic forces between electrified bodies obeyed an exact inverse square law. This idea can be traced back to Benjamin Franklin's observation that cork balls suspended in a charged cup were not drawn to either side, and the analogy Joseph Priestley pointed to: Isaac Newton's demonstration that a massive sphere exerted no attraction on masses inside it (Heilbron, 1979, p. 463).

The next year, in 1874, Maxwell acquired the unpublished electrical researches of Henry Cavendish. He found that a hundred years earlier, in 1772-73, Cavendish had tested the prediction of no interior electricity experimentally. Cavendish had concluded that the negative exponent in the force law could not differ from two by more than about 1/50 (Maxwell, 1879, pp. 111-112). Maxwell and his student, Donald MacAlister, created their own version of Cavendish's experiment, claiming a sensitivity of 1/21600. They published their experiment shortly before Maxwell's death, in his edition of *The Electrical Researches of Henry Cavendish* (1879, pp. 417-422), and in the posthumous second edition of Maxwell's *Treatise* (1881, pp. 77-82). Here Maxwell stated even more emphatically than in the first edition that a null result – this time Cavendish's – was "… a far more accurate verification of the law of force [than Coulomb's]" (p. 77).[1] He presented his own experiment as an improvement upon the accuracy of Cavendish's.

Why, given Maxwell's expressed confidence in Faraday's null demonstrations, did he bother repeating Cavendish's experiment?[2] Taking on board critiques of the experiments by Dorling (1974) and Laymon (1994), this paper traces the historical and conceptual re-orderings through which Maxwell aimed to secure Coulomb's law, and his motivations for doing so. It begins in §2 by examining the changing status of the inverse square law in Britain in the 1870s, concluding that it was only just becoming widely accepted. Yet, as will be discussed in §3, the law was an essential foundation for the electrical programme based on precision measurement and absolute units that Maxwell and William Thomson were attempting to establish (Smith & Wise, 1989, pp. 120-128, 237-276; Hunt, 2015). Despite the flawed logic of the mathematical tradition underpinning the null demonstration, the need to ensure acceptance of Coulomb's law drove Maxwell and Thomson's attempts to establish it by constructing an experimental tradition of null electrical measurement (§4), and arguing for the superiority of such a method (§5). By identifying Cavendish's as a null method, Maxwell mediated between the practices of telegraph engineers and the mathematical theory of electricity, through his developing 'doctrine of method' (§6).

---

[1] This statement can be attributed to Maxwell as it comes from within the first nine chapters of the second edition of the *Treatise*, which were completed and in press before his death; Niven in Maxwell (1881), p. xv.

[2] Maxwell said that he was "repeating" Cavendish's experiment, and I use his term throughout. While it might be possible to consider it as a "replication" of Cavendish's, fitting Radder's (1992) classification of reproducing the result of the experiment (the inverse square law) by a (not very) contemporary scientist, this would entail expanding the domain of "replication" to an experiment that was intended to affirm, rather than verify, the inverse square law (see also fn. 9).

## 2. The status of the inverse square law

In the years leading up to publication of Maxwell's *Treatise*, perceptions in Britain of the status of the inverse square law varied widely and were changing rapidly. Views ranged from William Thomson's that it was a mathematical truth (discussed later), to those of the prominent electrician, William Snow Harris, and the submarine cable engineer, Frederick Charles Webb, that it was not fundamental and held only in some circumstances (Webb, 1862, pp. 109-111; Harris, 1867, pp. 31-49; Thomson, 1872, pp. 24-25).[3] Since the 1840s two lines of reasoning had been evident. Harris led an inductive experimental one, based on Coulomb's and his own direct quantitative measurements. Thomson promoted a mathematical deductive argument based on potential theory and qualitative observations by numerous electricians of the absence of electricity inside conducting shells – evidence whose relevance was indirect and was not always made explicit. This was the line Maxwell took in 1873.

Harris and his followers pointed out that Coulomb had only ever published one experiment, comprising three data points, to support his conclusions (Coulomb, 1884). This left plenty of scope for measurements in different conditions to produce different laws. Between 1834 and 1839 Harris investigated high-tension static electricity (1834, 1836, 1839). He was one of the few to criticise Coulomb's experiments directly, seeking, "… by operating with large statical forces… to avoid many sources of error inseparable from the employment of very small quantities of electricity, such as those affecting the delicate balance used by Coulomb" (1839, p. 215). He concluded that Coulomb's law represented the composition of a fundamental direct inverse relation with distance, and changes to the distribution of electricity due to induction; it held only with suitable arrangements of conductors. His conclusion seems based as much on metaphysical reasoning that, "it is highly probable, if not morally certain, that every physical effect is in simple proportion to its cause" (1867, p. 209), as on his experimental measurements. Harris continued to promote this composition view until his death in 1867, and in his posthumous *Frictional Electricity* (1867). Webb followed him in suggesting that, "…. although the attractive force may … vary within certain limits, sensibly as some particular power of the distance, if increased beyond these limits, the ratio will begin to vary, and ultimately the attractive force will vary as some other power of the distance" (1862, p. 148).

However, Thomson had opposed Harris' interpretation since 1845, deeming his experiments incompetent (Smith & Wise, 1989, pp. 217, 246). Thomson assumed the inverse square law, and used his new method of electrical images to calculate the macroscopic force-distance law between Harris' conductors. By challenging the achievable accuracy of Harris' experiments, he was able to claim that his calculations agreed with Harris' observations. So, Harris' results supported rather than contradicted Coulomb's law (Thomson, 1872, pp. 24-25). In the same paper Thomson began promoting the mathematical argument for the null result as proof of the inverse square law (see §3).

Two series of textbooks provide evidence for the rising authority of the mathematical approach to electrical science among British electricians in the 1860s and 70s. They are by Edmund Atkinson, a physics teacher with a chemical background, and by Joseph D. Everett, a pupil of Thomson's who became Professor of Natural Philosophy at Queen's College, Belfast.[4] Atkinson's and Everett's books were translations of popular French textbooks by Adolphe Ganot and Augustin Privat-Deschanel respectively. The French market, mainly for medical students, provided a source of good textbooks in experimental physics at the time (Simon, 2015, pp. 28).

---

[3] Harris (1791-1867) is best remembered for his work on lightning conductors, especially on ships, for which he was knighted in 1847; James (2004). Webb (1828-1899) was involved in many cable projects including the successful Havana to Key West, and from Marseilles to Algiers, and the less successful early Atlantic cable; *Electrician* (1885), Yavetz (1993).

The earlier book, by Atkinson (following Ganot), asserted the inverse square law together with a description of Coulomb's experiment, giving his three data points (Atkinson, 1866, p. 559). He included a section, that Ganot had added in 1857, on Harris who, "has found that Coulomb's first law does not obtain in cases where the two bodies are charged with unequal quantities of electricity…" (Ganot, 1857, p. 538; Atkinson, 1866, p. 561). Atkinson next described experiments by Coulomb, Jean-Gustave Bourbouze,[5] and Jean-Baptiste Biot with closed spheres, and Faraday with open cylinders and cones, showing that electricity resides entirely on the surface of a conductor with none inside. At this point Ganot had made clear that these observations might be related to the inverse square law. He presented the observation as a consequence of the law: "En effet, en soumettant au calcul l'hypothèse des deux fluides, et en admettant qu'ils s'attirent mutuellement en raison *inverse du carré* de la distance et qu'ils repoussent leurs propres molecules suivant la meme loi, Poisson est arrive à la meme consequence que Coulomb sur la distribution de l'électricité libre dans les corps," (Ganot, 1856, p. 538, my emphasis).[6] Atkinson mistranslated this passage: "Admitting the hypothesis of two fluids, and that opposite electricities attract each other in the *inverse ratio* of their distances…" (1866, p. 564, my emphasis). Whether this was a typographic error, or whether it evidences Atkinson's confusion over the law of electrostatics, it was unfortunate given the contested status of the law in Britain.

In the 1870 edition, Atkinson had dropped the discussion of Harris and advanced no objections to Coulomb – evidence of the impact of Harris' death in 1867 on reducing promotion in Britain of the inductive experimental approach. However, his mistranslation of the inverse square law passage remained uncorrected in this and subsequent editions (Atkinson, 1870, p. 619).

Thus, by 1870, experiments showing that electricity resided only on the surface, and not inside, conductors, were deemed relevant to the laws of electrostatics, at least by some authors. But Everett's 1872 translation of Deschanel was the first British textbook to promote Thomson's deductive approach to proving the inverse square law. Everett alluded to Harris only obliquely when he claimed, "Many persons have adduced, as tending to overthrow Coulomb's law of inverse squares, experimental results which really confirm it" (Everett, 1872, p. 522). In this he was making the same claim that his mentor, William Thomson, had made in 1845 (see above).

---

[4] The market for such textbooks, and Atkinson's career and role are discussed by Simon (2015), especially pp. 76-86. For Everett's career, see Lees (2004)

[5] Bourbouze was *preparateur* at the Sorbonne and a teacher of experimental physics. His work is discussed by Blondel (1997)

[6] "Indeed, through calculations based on the hypothesis of two fluids, and that opposite electricities attract each other *inversely as the square* of the distance and repel like molecules according to the same law, Poisson arrived at the same conclusion as Coulomb on the distribution of free electricity on bodies" (my translation). A reviewer has pointed out that Ganot's treatment of Poisson here may be an anomaly among French textbook authors. Ganot was renowned for the exceptionally concise pedagogy with which he linked facts, theories and concepts. He had begun his career as a teacher of mathematics before moving into physics, in contrast to the majority of writers who started as chemists or medics; Simon (2015), especially pp. 68-72, 16-117. Further research would be needed to establish this point.

Following Deschanel (and Ganot and Atkinson), Everett described Coulomb, Biot, and Faraday's observations of the surface distribution of electricity (Privat-Deschanel, 1868, pp. 536-543). But he augmented Deschanel's description with three sections, the first being on technical limitations. The second detailed Faraday's (1843) experiments demonstrating electricity on the outside, but not the inside, of an (open) ice pail (see §4). The third, "No force within a conductor" described Faraday's (closed) 'cage' experiments (Faraday, 1838, p. 5). Where Deschanel had said nothing about a connection between the law of electrostatics and the surface distribution of electricity, Everett now stated categorically the relevance of Faraday's 'cage' experiment:

> The fact that electricity resides only on the external surface of a conductor, combined with the fact that there is no electrical force in the space inclosed [sic] by this surface, *affords a rigorous proof of the law of inverse squares…. Now it admits of proof, and is well known to mathematicians*, that a uniform spherical shell exerts no attraction at any point of the interior space, if the law of attraction be that of inverse squares, and that *the internal attraction does not vanish for any other law,* (Everett, 1872, p. 521-22, my emphasis).

Here Everett did two things: he asserted that internal attraction does not vanish for any law other than an inverse square (discussed in §3), and; he implied that mathematicians were the only group with a proper knowledge of electrostatics. This passage, when combined with the evidence above that Atkinson was retreating from Harris' claims, shows the beginnings of the ascendancy of mathematical deduction over experimental induction in textbook discourse by the early 1870s. Even Webb, who was not a mathematician, concluded in 1868, "problems of electrostatics in their generality must belong to a very high class of mathematics," (1868, p. 325).

In the deductive approach, non-mathematicians were frequently expected to take the inverse square law on trust. Thus Everett, in his *Elementary Textbook of Physics* (not his translation of Deschanel), simply asserted the law, citing no evidence (1878, p. 252). Here he went even further than Fleeming Jenkin, Maxwell and Thomson's associate on the British Association Committee for Electrical Standards, whose earlier *Electricity and Magnetism* cited only unspecified experiments (1873, p. 95).

This investigation of some common textbooks suggests that by the early 1870s, when Maxwell wrote the *Treatise*, both the inverse square law, and the dominance of the mathematical approach to electrical science, were becoming established in Britain, the fate of the former apparently intimately connected with the progress of the latter. By around 1877 when he repeated Cavendish's experiment, the law could be asserted without question. The mathematical logic of this experiment is examined in the next section, en route to establishing the central importance of the inverse square law in the deductive approach in §4.

## 3. Mathematical deduction and the logic of the null experiment

In a review of Jenkin's 1873 textbook, widely attributed to Maxwell, he made a distinction between a science of "sparks and shocks which are seen and felt," and a science of "currents and resistances to be measured and calculated" (*Nature*, 1873, p. 42). Thomson had argued the latter view since 1845. Despite differences over the underlying processes of electromagnetic phenomena, he and Maxwell were able to agree about the primacy of measurement and mathematical law and to promote these in what we might call their "electrical programme". In a series of moves that aligned with their advocacy of energy physics more generally, they operationalised their beliefs in such activities as the British Association Committee on Electrical Standards and the associated development, by Thomson, of a variety of precision instruments (Smith, 1998). They worked to unify electrical science and practice with better-established and mathematised branches of physics through the derivation and implementation of "absolute" units of measurement. These related electrical quantities to mechanical force or units of length, mass, and time (Smith & Wise, 1989, pp. 684-94; Hunt, 2015; Mitchell, in press). For both men, the inverse square law underlay the mathematical treatment of electrostatic problems.

The early priority Thomson ascribed to the law is shown in his paper on "The Mathematical Theory of Electricity in Equilibrium", first published in 1845 and re-published with additional notes in 1854 (Thomson, 1872), discussed in depth by Buchwald (1977). Coulomb's law must be true, Thomson believed, because mathematics determined its form: "…[Coulomb] has thus arrived by direct measurement at the law, which we know by a mathematical demonstration, founded upon independent experiments, to be the rigorous law of nature for electrical action," (pp. 24-25). He referenced Robert Murphy (1833), and John Pratt (1836) for the mathematical demonstration, but did not specify the "independent experiments". In 1854, Thomson added a note: "Cavendish demonstrates mathematically that if the law of force be any other than the inverse square of the distance, electricity could not rest in equilibrium on the surface of a conductor. But experiment has shown that electricity does rest at the surface of a conductor. Hence the law of force must be the inverse square of the distance," (p. 24 fn). Here he specified the relevant class of experiments for the first time, but remained silent about who had performed them. He was unaware then that Cavendish had actually performed the experiment. However, as mentioned in §2 above, observations by electricians such as Coulomb, Biot, Bourbouze and Faraday, over a range of conductor shapes, tended to this result.

Cavendish's mathematical demonstration utilised a one-fluid model of electricity and implicitly tested Priestley's analogy between electric and gravitational force. Cavendish assumed an exact inverse square law, and showed the electricity on a closed spherical conductor was confined to the shallowest possible depth on the surface, with no electricity inside (1771, pp. 592-4). According to Jungnickel and McCormmach, "so compelling was the example of the law of gravitation that Cavendish did not consider the possibility that the distance dependency of the electric force could be anything but some inverse power of the distance" (1996, p. 184)[7]. However, being cautious, he did entertain the possibility of inverse powers other than two. He discussed qualitatively the consequences of powers between one and two, and two and three, and reasoned that either case would lead to observable interior electrification (1771, pp. 594-5). Cavendish published this theoretical result in 1771, but until Harris, and then Maxwell, examined his papers in the 1860s and 70s, no one knew he had also tested it experimentally, concluding, "the electric attraction and repulsion must be inversely as the square of the distance" (Harris, 1867, p. 45; Maxwell, 1879, p. 110).

---

[7] Cavendish's electric fluid comprised finite-sized particles. He wrote frequently of them being compressed as tightly together as possible. So he knew, though he did not articulate, that there was a minimum distance to which a simple power law would apply. His demonstration really applies only at distances greater than this minimum, where he seems to have considered a power law unproblematic.

Consider a charged spherical conducting shell. By symmetry, and since the electricity can move around on the conductor, it will distribute itself uniformly over the surface. The situation is, then, similar to that of a particle inside a massive shell, considered by Newton in proposition 70 of Book 1 of the *Principia*, to which Cavendish referred (1771, p. 592). In modern terms, consider the electric force at any interior point, P, by drawing a cone of infinitesimal solid angle through P that cuts the sphere on both near and far sides (Figure 1). For an inverse square law, the force at P is $\frac{\rho dS_n}{r_n^2} - \frac{\rho dS_f}{r_f^2}$ where $\rho$ is the surface density of electricity, $dS_n, dS_f$ are the infinitesimal surface areas of the sphere cut by the cone on near and far sides, and $r_n, r_f$ are the distances from the point to the surface of the sphere.

FIGURE 1

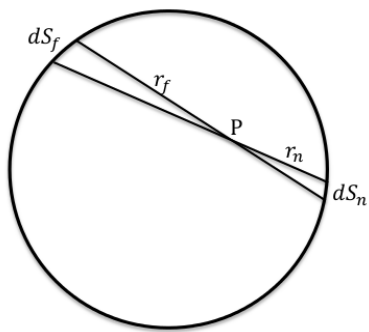

*Figure 1. The geometry of an inverse square force on a point inside a spherical conducting shell.*

By geometry, the infinitesimal areas $dS_n, dS_f$ increase as the square of $r_n, r_f$, so write them as $dS_n = \alpha r_n^2$ and $dS_f = \alpha r_f^2$. Hence the force at P becomes $\frac{\rho \alpha r_n^2}{r_n^2} - \frac{\rho \alpha r_f^2}{r_f^2} = 0$, since the $r_n^2$, $r_f^2$ terms cancel out. This result holds whatever the direction of the cone. Thus, for an inverse square microscopic force the total resultant force inside the shell is everywhere zero. If the microscopic force obeys a power law that differs from inverse square by a small amount, the resultant will be directed towards or away from the centre. Then, argued Cavendish, suppose a globe is placed inside the outer shell, connected to it by a wire, and consider the free electric fluid in the wire. If the inverse power differs from two, the fluid will move towards (or away from) the centre, leaving the inner globe with excess (or deprived of) fluid. Either way, unless the power is exactly two, the inner globe will be electrified.

Thomson and Maxwell interpreted Cavendish's demonstration and, later, experiment, using potential theory: if the microscopic law of force between two charges is inverse square, then all the electrification of a conducting shell resides at the surface with none inside. The zero resultant force inside the shell indicates that the electric potential is constant throughout the interior.

The experiment involved arranging an insulated conducting globe inside a concentric hollow conducting sphere comprised of two separable hemispheres (see Figure 2). Initially, a wire connected the globe and sphere. The apparatus was electrified from a Leyden jar, then the connecting wire was removed and the outer sphere discharged to earth. Testing the electrification of the inner globe with an electrometer showed it to be nil.

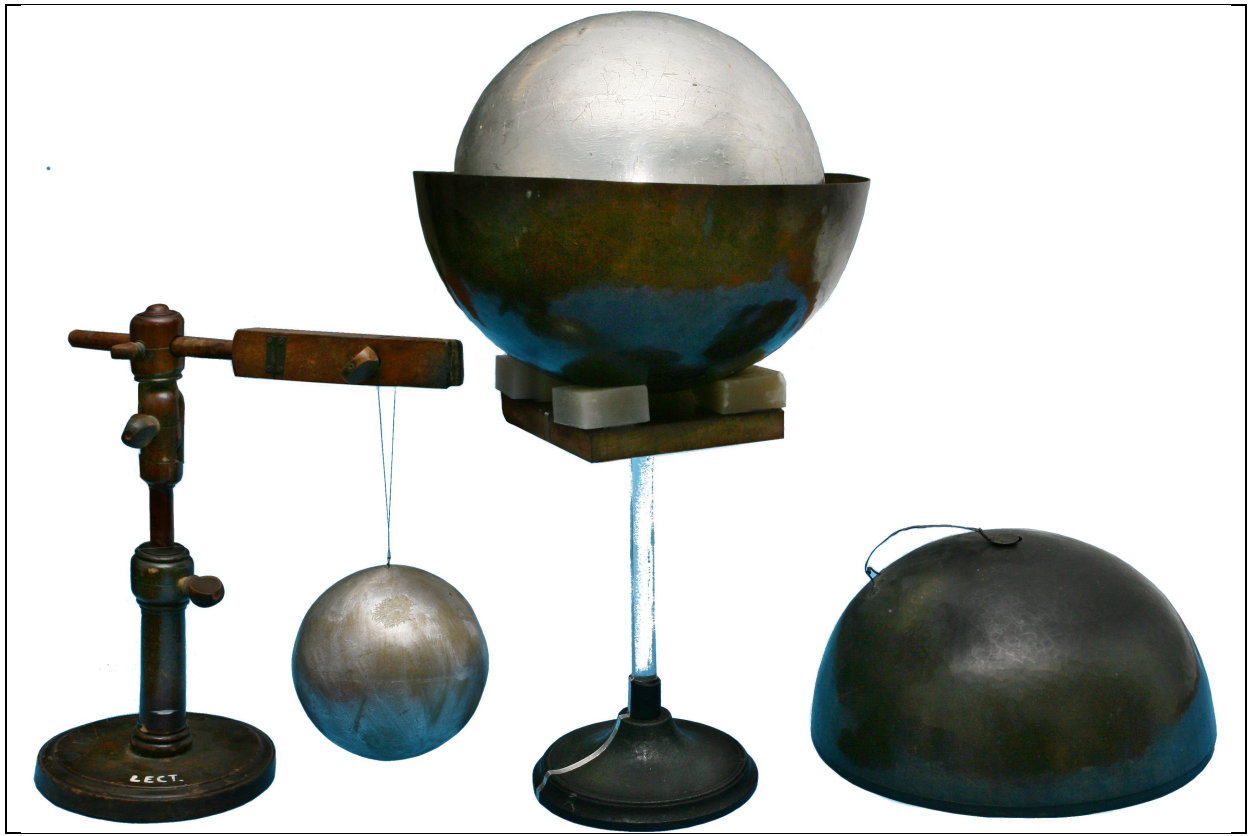

*Figure 2. Maxwell and MacAlister's apparatus. The additional ball hanging from the retort stand has traditionally been considered part of the apparatus, but does not correspond to the brass ball for testing sensitivity of Maxwell's description. Photograph courtesy of the Cavendish Laboratory, Cambridge.*

This indirect null method has formed the basis of increasingly precise demonstrations of Coulomb's law ever since, notably by Plimpton and Lawton (1936), Bartlett et al. (1970), Williams et al. (1971) and Crandall (1983). However, Dorling (1973, 1974) and Laymon (1994) have critiqued it on epistemic grounds.

Dorling took a Bayesian approach, in which the probability of an initial proposition such as the inverse square law is updated in the light of new evidence such as the null result. He concluded that Cavendish's method of arriving at the law was rational and an example of "demonstrative induction" in which a universal generalisation (the law) is deduced from a particular instance of itself (Cavendish's experiment). But it relied on an initial assumption – that the microscopic force was an inverse power of the distance – which he claims would have been implausible to Cavendish and his eighteenth-century contemporaries. If the force obeyed a distance law at all (which those who thought atmospheres of electric fluid surrounded electrified bodies had no reason to believe), then a power law was over-restrictive; it applied neither to the dominant Boscovichean action-at-a-distance theory, nor, in fact, to Cavendish's own electric fluid whose particles could not get closer together than two particle radii (Maxwell, 1881, p. 82; Dorling, 1974, p. 334; Heilbron, 1979, pp. 376, 426).[8] Maxwell, in turn, seized this opportunity to boost both the plausibility of a power law, and his own views of electricity, by arguing, "so little like ordinary matter as electricity appears to be," conceptualising electricity as particles was inappropriate (1879, p. 422). This attempt prioritised mathematical reasoning over easy physical conception and Dorling judged that it was unsuccessful in establishing the validity of the initial assumption (1974, p. 336).

Laymon (1994) problematised Dorling's attribution of demonstrative induction by pointing out that both Cavendish and Maxwell had assumed the inverse square law when assessing the sensitivity of their experiments (see §5). Laymon included twentieth century experiments in positing a developmental process in which successive experimenters have assumed, in their sensitivity analysis, the inverse square law to at least the precision of their predecessors.[9]

A further problem with Maxwell's and Cavendish's arguments has seldom been noted, though it is implicit in Dorling's account: although an inverse square law predicts, with absolute certainty, that all the electricity resides on the outside of a closed conductor, the converse is not necessarily true. Maxwell was well aware, in general terms, of this problem of reverse inference, and his draft account of Cavendish's methodology raised it.

> It is obvious that the mere agreement between experiment and the deductions from an hypothesis cannot prove that hypothesis to be true unless it can also be shown that no other hypothesis will agree with the experiments … we must form such an exhaustive classification of hypotheses as to enable us from the result of an experiment to conclude that all hypotheses except those belonging to a certain class must be wrong (Harman, 2002, pp. 541-542).

Maxwell took a mathematical approach to demonstrating that no other hypothesis would agree with the null result. Already, in 1873, he countered mathematically any who still believed the law might vary in different circumstances by stating that an inverse square was, "the only law of force which satisfies the condition that the *potential* within a uniform spherical shell is zero," (1873a, p. 76, my emphasis). As noted above, Thomson had referenced the Cambridge-educated mathematicians Robert Murphy (1833) and John Pratt (1836) when making a similar claim in 1845.[10] Maxwell followed this tradition by referencing Pratt. But although he cited Pratt, Maxwell's detailed demonstration was due to his friend, Peter Guthrie Tait, outlined in a postcard dated 5 June 1871 (Harman, 1995, p. 650).

---

[8] Dorling's *logical* argument about what should have been implausible contrasts with Jungnickel and McCormmach's (1996), p. 184, *historical* claim about Cavendish's supposed Newtonian commitments. Cavendish did not articulate the problem of a minimum distance for the power law (see fn. 7).

[9] The issue here differs from the well-known "experimenters' regress" found in the replication literature in which the status of truth claims is at stake, e.g. Radder (1992), Collins (1985). In this series of inverse square law experiments few people, since the early 1870s, have doubted the law. The experiments have been about how accurately the power could be measured, and their purpose in relation to the truth of the law has been to affirm it rather than to test it. The process is closer to Chang's epistemic iteration, being an iteration towards experimental precision, although it differs from his examples in iterating towards a pre-determined outcome, a characteristic Chang ascribes to mathematical iteration (2007), p. 18.

[10] Robert Murphy (1807-1843) graduated from Cambridge as third wrangler in 1829, becoming a fellow of Caius College. In 1833 he published *Elementary Principles of the Theories of Electricity, Heat, and Molecular Actions*, drawing on the work of Ampere, Poisson, and Green: Crilly (2004a); Robertson and O'Connor (2015). John Henry Pratt (1809-1871) was Cambridge third wrangler in 1833. He studied the figure of the Earth, taking Laplace's work as his starting point; Pratt (1836). His later work on the mass of the Himalayas led to the hypothesis of uniform depth of compensation, or "Pratt isostasy"; McConnell (2004).

Both Pratt's and Tait's demonstrations suffered from the same fundamental flaw if they were to be realised experimentally: they ignored the quantifiers at a crucial step in the argument. This is very clearly seen in Maxwell's 1873 reproduction of Tait's demonstration. Assuming a radially symmetric law, Maxwell began by defining the potential function due to a unit of electricity at a distance $r$ as $f(r)/r$. Then he derived the potential $V$ at a point within a spherical shell of radius $a$ at a distance $p$ from the centre by integrating over the surface of the shell. He obtained $V = 2\pi\sigma\{f(a+p) + f(a-p)\}$, where $\sigma$ is the surface density of electricity. Using the experimental observation that there is no electricity inside the shell, and hence that $V$ is constant, he differentiated $V$ with respect to $p$ to obtain $f'(r)$ in the first of the crucial, but problematic, three lines from which he derived the inverse square law:

$$0 = f'(a+p) - f'(a-p)$$

Since *a* and *p* are independent,

$$f'(r) = C, \text{a constant} \quad (1873a, p. 76).$$

Integrating $f'(r)$ to obtain $f(r)$, substituting into the expression for the potential function, and differentiating the potential function to give the electric force, gave the force as proportional to $\frac{1}{r^2}$. The problem is that the third line above is true only if the first line applies *for all a* and *p<a.* This can be seen by starting with the first line, which requires that $f'(a-p) = f'(a+p)$. Then, if $a$ is held constant, any function $f'(r)$ which is symmetric about $a$ satisfies the requirement when $r = a - p, r = a + p$ as $p$ varies independently. So $f'(r) = C$ works, but so does $f'(r) = \cos(r-a)$ or $f'(r) = (r-a)^2$ (or a multitude of other even functions). If $a$ is now allowed to vary as well as $p$, then the only function that still satisfies the requirement is $f'(r) = C$. Maxwell stated in the second line above that $a$ and $p$ are independent but forgot, or did not realise, that he had to allow $a$ as well as $p$ to vary in order for his conclusion to hold.

Thus, if applied to a single experiment, with a sphere of fixed radius $a$, then all the first line of Maxwell's demonstration tells us is that $f'(r)$ is symmetrical about the point $a$. To prove the third line, and hence the inverse square law, experiments need to be performed with spheres of many different radii. But the lack of quantifiers meant that this was not obvious to anyone reading the demonstration without critical attention.

Six years later, in 1879, Maxwell dropped Tait's demonstration when reporting his repetition of Cavendish's experiment. Instead he asserted the result as fact, calling on the authority of Pierre-Simon Laplace who "gave the first direct demonstration that no function of the distance except the inverse square satisfies the condition that a uniform spherical shell exerts no force on a particle within it" (1879, p. 422; 1881, p. 82). Maxwell's source here was probably Isaac Todhunter's account of Laplace, published in 1873. Todhunter's demonstration, and Laplace's original, are similar to Maxwell's, but differ crucially in stating the quantifiers (Laplace, 1799, p. 143; Todhunter, 1873, pp. 181-182). However, they do not spell out the experimental implications. Dorling has described Maxwell as "disingenuous" for glossing over the essence of Laplace's mathematical demonstration – shells of all possible radii would need to produce a "no force" result in order for us to infer an inverse square law (1974, p. 336).

Although Dorling is mathematically correct, he is unduly harsh, for he has viewed Maxwell's work in isolation. Maxwell's reference to Pratt, even while using Tait, and the similarity of his general argument to Thomson's of 1845, suggests he was following Thomson's lead. Murphy, Pratt, and Tait all neglected to state quantifiers (Murphy, 1833, p. 41; Pratt, 1836, pp. 142-143; Harman, 1995, p. 650). In 1879 Maxwell was uncritically following in an established tradition, rather than Laplace whom he cited, and the problem arose only when the demonstration was applied to a single experiment. Nor was his experiment solitary. Some use of induction is clearly reasonable; if Maxwell's is considered as merely the most accurate of the wide range of null observations, with different sizes and shapes of shell, described in the textbooks, then Dorling's criticism loses much of its force.

Having established the mathematical logic – albeit a flawed one – of the null experiment, we now examine the need to ensure uncritical acceptance of Coulomb's law, and Maxwell, Thomson and Everett's attempts to establish it by constructing an experimental tradition of null tests.

## 4. Inventing a tradition of testing the inverse square law by null experiments

In his seminal history of *Electricity in the 17th and 18th centuries*, John Heilbron, suggests the analogy drawn by Priestley between electric and gravitational force was so powerful that most electricians believed it, despite its manifold problems. But an open cup, as used by Franklin and Priestley in their experiments with corks, is not a closed sphere. Furthermore, electricity can redistribute itself on conductors in a way that gravitational masses cannot. And, as we have seen, arguing from a closed shell to zero force is not the same as arguing the converse. Authors such as Tiberio Cavallo, William Nicholson, Martinus van Marum, and Girolamo Adelasio pointed out these problems in the 1780s (Heilbron, 1979, p. 463).

Where Cavendish had contemplated other power laws, Maxwell was less cautious. He was heir to the potential theories of Laplace, Carl Friedrich Gauss, George Green and Thomson, whose technical power provided a strong incentive to uphold the inverse square law. In the 1840s, Thomson had transposed the idea of a 'potential function' from an analytic tool to a mathematical description of a physical state whose gradient provided electric force (Smith & Wise, 1989, p. 207). This enabled him to avoid speculation about the microscopic nature of electricity. Instead, drawing also on Green's (1828) paper on the uses of potential theory, Thomson developed methods for treating electrostatic theory observationally – using differential equations for macroscopic quantities that could be measured and interpreted using potentials (Buchwald, 1977). Thomson and Maxwell based their electrical programme, discussed in §3, on this macroscopic, potential, approach. Hunt has characterised the approach as "engineering" and pointed out its persuasive value to telegraph engineers (2015, pp. 305, 328).

However, experimental proof of an inverse square law is a *necessary* condition for potential theory as developed by Thomson to be of any use in electrostatics. By the time Maxwell was preparing his *Treatise,* evidence for the absence of electricity inside a conducting container had been augmented by the experiments of Coulomb, Bourbouze, Biot, and Faraday, mentioned in most textbooks. Like Everett in his 1872 textbook, and as we saw in §3, Maxwell argued that the absence of interior electricity demonstrated the validity of the inverse square law (1873a, p. 75). While Everett had described Faraday's (1843) experiments with an *open* ice pail, he based his claim of a rigorous proof of the inverse square law on Faraday's (1838) *closed* cage experiments (§2). In contrast, the only evidence Maxwell cited was Faraday's *open* ice pail experiments. These were the experiments he rated as, "far more conclusive than any measurements of electrical forces" (1873a, p. 75).

To bolster the observational evidence, Maxwell not only shifted the evidential context of Faraday's experiments, but also distorted their description. Inspection of Faraday's 1843 paper shows he made the experiments to demonstrate his own ideas about electrostatic induction rather than to investigate the law of force. Faraday's diagrams (and Everett's reproduction of them) clearly showed his conducting shells as cylinders, closed at the bottom but wide open at the top. In the text, he described them as "pails". In his own text Maxwell frequently referred to the containers as closed. In his diagram, without noting the change, he showed the cylinder as fully closed (and with rounded ends to look more nearly spherical) (Figure 3). This step was necessary for his argument, since the prediction of potential theory applies only to *closed* conductors. By closing the shell in text and diagram, Maxwell betrayed his awareness that, mathematically, Faraday's experiments provided no evidence for the inverse square law.

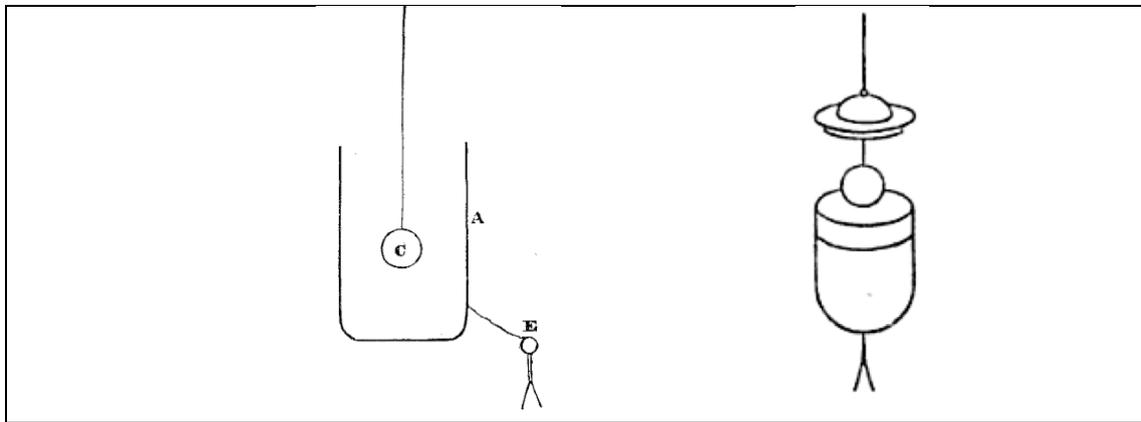

*Figure 3. Faraday's diagram of his apparatus (left) (1843, p. 200), and Maxwell's diagram of Faraday's apparatus (right) (1873a, p. 31).*

That Maxwell arranged for the small access hole in his own experimental shell to be covered while the conductors were charged, provides further evidence for his recognition of the mathematical importance of a closed shell (Figure 4 below).

Thomson was even more emphatic than Maxwell about Faraday's role in demonstrating the absence of electricity inside a hollow electrified conductor. Speaking in 1876, before he knew of Cavendish's experiment, he held that,
> The other main method of experimenting in connection with measurement … is illustrated also by Cavendish's writings, that is the seeking for a zero…. It was left for Faraday to make with accuracy the concluding experiment which crowned Cavendish's theory. Faraday [1843] found by the most thoroughly searching investigation that the electrical force … was zero…. Therefore the law of force varies with the inverse square of the distance. This result was obtained with far less searching accuracy by Coulomb and Robinson[sic], because their method did not admit of the same searching accuracy [sic] (1876, p. 241).

Despite Thomson's claims of accuracy, another look at Faraday's paper shows he had used a "delicate" gold leaf electroscope to give a visual indication of the varying electrification induced on the outer cylinder when a charged ball was introduced. "If C [the ball] be made to touch the bottom of A [the cylinder] … C, upon being withdrawn and examined, is found perfectly discharged." Notably, he gave no indication of the means by which he had examined the electrification of the ball, finding it to be zero. He clearly thought this unnecessary, for he continued, "These are all well-known and recognised actions" (Faraday, 1843, p. 201).

Thomson and Maxwell defended the inverse square law vigorously, but their defences were always predicated on the general knowledge that there is, indeed, no electricity inside a closed conductor. The power of the mathematics available to them, and their investment in an electrical programme based on it, gave them the confidence to overlook or distort the logical and mensurational weakness of Faraday's experimental evidence in asserting the validity of the law.

Morus (1993) has commented on the importance to Thomson and Maxwell of reconstructing Faraday's researches into a genealogy for their own electrical programme – a reconstruction that is well evidenced in their treatment of Faraday's "null" experiments. Falconer (2015) advances a similar explanation for why Maxwell devoted five years of his life to editing Cavendish's researches, pointing out that this genealogy was British, aimed at a British audience. Repetition of Cavendish's experiment was a part of this effort that became particularly important if Maxwell had realised the weakness of using Faraday's work as a basis. Even if he had not, adding Cavendish to the genealogy could only strengthen the case.

## 5. Accuracy and the value of Maxwell's null experiment

The previous quotation from Thomson illustrates the argument that Maxwell and he constructed for the superiority of the null method. It was superior because it was intrinsically more accurate than Coulomb's method. Precision measurement, and associated claims of accuracy, went hand in hand with mathematical analysis in Thomson and Maxwell's electrical programme. This section examines the arguments for the accuracy of the null experiment.

Thomson attributed accurate knowledge of the inverse square law to the care of Faraday's "thoroughly searching investigation", and the power of Cavendish's null method. But what was wrong with Coulomb and John Robison's experiments? Maxwell's remarks on the value of Faraday's null experiment were preceded by his objection to Coulomb's method.

> The results, … which we derive from such experiments must be regarded as affected by an error depending on the probable error of each experiment, and unless the skill of the operator be very great, the probable error of an experiment with the torsion-balance is considerable (1873a, p. 75).

Coulomb himself gave no indication of the likely accuracy of his experiments (1884). Heering's (1992) reconstruction suggests they were unlikely, in practice, to have shown more than that the inverse power lay between one and three, while Martínez's (2006) gave results nearer to two. But both reconstructions show the exponent varying as the distance between the charged balls increases. In this sense they are more in line with Harris' results than with Coulomb's claims.

Turning from Coulomb, where did Maxwell and Thomson think the "searching accuracy" of Faraday's experiments lay? Maxwell suggested it was in the sensitivity of the instruments used to detect electrification:

> … The methods of detecting the electrification of a body are so delicate that a millionth part of the original electrification of B [the charged ball] could be observed if it existed. No experiments involving the direct measurement of forces can be brought to such a degree of accuracy (1873a, p. 75).

However, as we saw in §4, Faraday gave no indication of the accuracy of his observations, nor sufficient detail for readers to estimate it. For Maxwell's rhetorical purposes, the superiority of Faraday's evidence over Coulomb's lay not in the actual delicacy or sensitivity of the instruments, but in his reputation in Britain as an experimenter.

The distinction Maxwell made in the quotation above, between 'detecting' electrification and 'measuring' a force, emerges as key to his argument for the superiority of the null method. We see this again four years later, in his commendation of Cavendish's experiment in a draft paper for the Cambridge Philosophical Society,

> Cavendish thus established the law of electrical repulsion by an experiment in which the thing to be observed was the absence of charge on an insulated conductor. *No actual measurement of force was required*. No better method of testing the accuracy of the received law of force has ever been devised… (Harman, 2002, p. 539, my emphasis).

Unlike Maxwell, Thomson called both methods "measurement" but he did distinguish them operationally. "In physical science generally, measurement involves one or other of two methods, a method of adjustment to a zero, or a what is called a *null* method, and again, a method of measuring some continuously varying quantity," (1876, p. 239, Thomson's emphasis). He credited Robison and Coulomb with establishing the inverse square law by measuring some continuously varying quantity, but Cavendish and Faraday with proving it accurately using a null method. The logical structure Thomson outlined was radically different; Robison and Coulomb's method was inductive, whereas Cavendish made a deductive prediction, which Faraday confirmed.

Maxwell and Thomson explicitly located the accuracy of Faraday's and Cavendish's null results in the instruments used, the care taken by the experimenter, and the claimed mathematical logic of the null method. In practice they used the reputation of the experimenter as a proxy for actual care, as seen when comparing their accounts with Faraday's own.[11] They did not articulate why the null method was better operationally, but by inference from Maxwell's criticism of Coulomb, it relieved the experimenter from some necessary care because only one observation was needed, rather than several. Moreover, that observation was of the absence of movement of the instrument's indicator. Many of the measuring instruments that Maxwell described in the *Treatise* relied on balancing electric or magnetic force against a restoring force, often from a torsion wire. When the electrical force was applied or changed, the indicator – pointer, suspended mirror, or fiducial hair – oscillated before settling down to the new equilibrium position corresponding to the measurement. Maxwell devoted pages of the *Treatise* to mathematical and manipulative techniques for reducing the time spent in determining accurately the new position for a variety of instruments (Maxwell, 1873b, pp. 336-351; see also Gooday, 2004, pp. 141-148).[12] Methods that produced no deflection, and hence no oscillation, might reduce both the time and manipulative skill required, especially if only one observation was taken.

---

[11] Gooday (2004) e.g., pp. 20-23, discusses the importance of reputation in engendering trust in results, interpretations and instruments.

Consideration of the improvements Maxwell made to Cavendish's experiment, and a comparison of their sensitivity analyses throws further light on his view of the burden of accuracy of the method and where care was necessary. He and his student Donald MacAlister improved on Cavendish's apparatus in two major ways.[13] First, where Cavendish had devised a hinged apparatus (AbcDCB in Figure 4) to remove the outer hemispheres and allow access to the inner globe for testing, Maxwell left the outer hemispheres in place throughout. This shielded the inner globe from possible electrical disturbances. He argued that it also removed one of the chief sources of error in Cavendish's experiment, leakage of electricity from the inner globe to the bench, as the globe now rested on insulating supports on the inside of the sphere. Thus "the potentials of the globe and sphere remained sensibly equal," so electricity was not attracted away from the globe (Maxwell, 1879, p. 417).

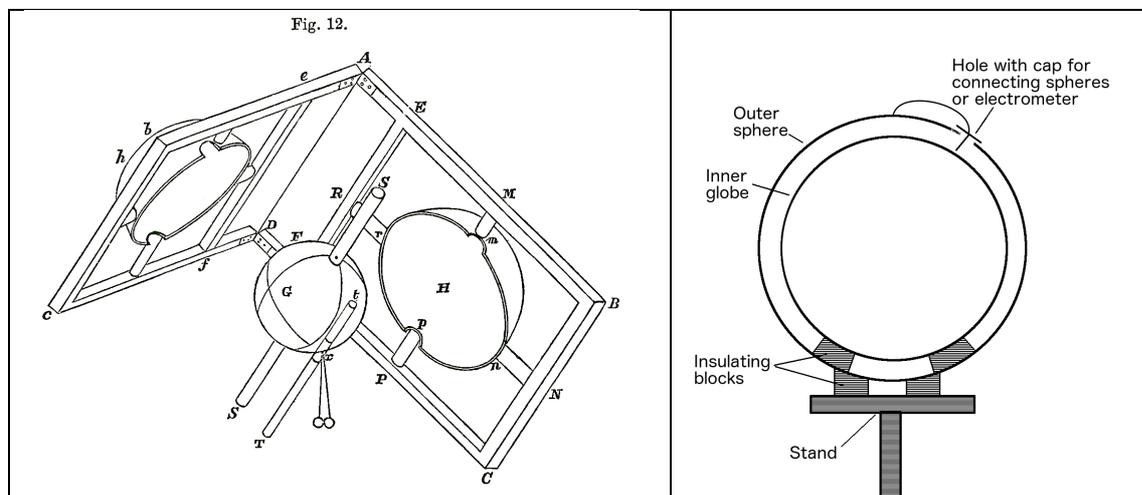

*Figure 4. Left: diagram of Cavendish's apparatus showing the hinged frame (AbcDCB), which supported the outer hemispheres (hH), the inner globe (G) supported by an insulator (SS) which stood on the bench, and the pith ball electroscope used for testing (Tt) (Maxwell, 1879, p. 106). Right: diagram of Maxwell and MacAlister's apparatus, showing the inner globe supported on insulating blocks within the outer sphere, which remains closed throughout except for a small capped hole (drawn from the original apparatus shown in Figure 2).*

Second, Maxwell tested for charge using Thomson's recently invented quadrant electrometer, which was far more sensitive than the pith ball electroscope available to Cavendish. Having assessed the sensitivity of the electrometer, as discussed below, Maxwell and MacAlister estimated that if the power in the law of repulsion were $-(2+n)$, then $n$ could be no greater

---

[12] The specific examples Maxwell gave are of galvanometers, but the principles of his analysis applied to electrostatic instruments also. I am grateful to Richard Staley and Daniel Mitchell for reminding me of this passage in Maxwell's *Treatise.*

[13] Donald MacAlister (1854-1934) was senior wrangler in 1877. He subsequently studied medicine, was President of the General Medical Council, and Principal, then Chancellor, of Glasgow University; Crilly (2004b).

than ±1/21600 – a 400-fold improvement on Cavendish's ±1/50 (Maxwell, 1879, pp. 112, 419; 1881, pp. 77-81).

As Laymon's account of the iterative series of null demonstrations implies, the null method displaces the burden of accuracy from the observation of the null result, to the measurement of the sensitivity of the detecting instruments. Mitchell (2010, pp. 177-8) has made a similar point in his account of Gabriel Lippmann's use of null experiments. Maxwell's account did not acknowledge such displacement explicitly, and despite the greater precision, his sensitivity analysis was more cursory than Cavendish's. Cavendish noted that the null position (i.e. with the pith balls collapsed) was the least sensitive manner of using his electroscope. He repeated his experiment with the electroscope in the more sensitive electrified position (Maxwell, 1879, pp. 108-110). Cavendish established the minimum detectable electrification of the globe by charging it from a condenser whose capacity was compared with that of a trial plate. Calculation of the capacity of the trial plate ultimately assumed the inverse square law (Laymon, 1994, pp. 43-45). The chain of reasoning was long and it is possible Cavendish was unaware of the issue.

Maxwell's description of his own sensitivity analysis relied on his readers' knowledge of the instruments he was using. Unlike Cavendish, he had a mathematical theory (based on the inverse square law) of how his electrometer worked. This enabled him to derive the limits of precision while always using the electrometer in the same, uncharged, mode. This was the customary mode of use even though, like Cavendish's electroscope, it was not the most sensitive possible. Maxwell showed that his quadrant electrometer could detect electrification on the outer sphere when that was charged, by induction from a small brass ball 60 cm away, to 1/486 of its previous electrification. The induced charge on the outer sphere gave an electrometer reading $D$. Maxwell asserted,

> The negative charge of the brass ball was *about* 1/54 of the original charge of the shell, and the positive charge induced by the ball when the shell was put to earth was *about* 1/9 of that of the ball. Hence when the ball was put to earth the potential of the shell, as indicated by the electrometer, was *about* 1/486 of its original potential, (Maxwell, 1879, p. 418; 1881, p. 79, my emphasis).

He did not state how he arrived at these estimates, but Laymon suggests he calculated them using Thomson's method of images, itself based on the inverse square law (1994, pp. 35-36).

The on-going analysis reveals more about Maxwell's lack of focus on measuring the sensitivity of his detection instrument. He continued,

> … let $d$, be the largest deflexion which could escape observation in the first part of the experiment [the null detection]. Then we know that the potential of the [inner] globe at the end of the first part of the experiment cannot differ from zero by more than
> $$\pm \frac{1}{486} \frac{d}{D} V$$
> where $V$ is the potential of the shell when first charged… Now, even in a rough experiment, $D$ was certainly more than $300d$. In fact no sensible value of $d$ was ever observed… (1879, pp. 418-419; 1881, p. 79).

This calculation assumed Maxwell's readers were familiar with the linear response of a quadrant electrometer at small measurement potentials, as analysed in his *Treatise* (1873a, pp. 271-274). In a quadrant electrometer, an arc-shaped conductor moves within a cylinder divided into four quadrants. The sensitivity of the electrometer is proportional to the potential of the moving conductor; the linear response depends crucially on maintaining this at a much higher potential than that of the quadrants. Maxwell gave his readers no indication of the potential used to ensure the linear response of his electrometer. Nor did he do more than a "rough experiment," to determine the largest deflection that could escape observation, and the estimate of "about" 1/486 relied on two preceding approximations.

These passages suggest that while Maxwell partially recognised the displacement of the burden of accuracy to the measurement of sensitivity, he did not lay it on the experimenter's manipulative or observational care in taking a reading. Instead, he attributed accuracy to the null form of the experiment, its careful design that reduced anticipated errors, and to the reputed sensitivity of Thomson's quadrant electrometer. This typifies his more general approach to promoting accuracy through pre-empting error beforehand rather than analysing it afterwards, as identified by Gooday (2004, p. 76). Despite the careful design, the lack of reported care in conducting the experiment suggests that Maxwell's aim in repeating the experiment was not purely metrological, to push it to the limits of precision.

Perhaps the experiment also served a rhetorical purpose, to promote to a community already convinced of the inverse square law, the precision that could be achieved using Thomson's instruments and null methods? Viewed in this light, Maxwell's disavowal of manipulative care was an advantage; with appropriate design and sensitive instruments, even operators whose skill, in contrast to Coulomb's, was not "very great" (1873a, p. 75) might achieve robust results. They could make observations rather than measurements, and fewer of them with consequent reduced scope for errors to accumulate. Although precision was estimated, it was trumped by careful prior design, promoting the mathematical theory on which the experiment was based. If this is the case, null methods played a role in the subordination of experimental to mathematical physics of Thomson and Maxwell's electrical programme.

This section has suggested that asserting the superior accuracy of the null method, and stressing the use of sophisticated instruments, promoted its acceptance, and diverted attention from the underpinning logic. Conversely, stressing the accuracy with which a null method demonstrated a law that was, by now, generally accepted, helped to secure both the method and the instruments necessary for its success. The strategy was one of mutual grounding. In this context, a robust sensitivity analysis was unnecessary. The next section examines the wider context around null methods within which Maxwell deemed these strategies desirable and likely to be successful.

### 6.   The wider role of null methods
In 1871 Maxwell had become the first Professor of *Experimental* Physics at Cambridge. His role was to establish experiment in a university hitherto dominated by mathematics, and to head the new Cavendish Laboratory, funded by the Duke of Devonshire to meet Britain's perceived need for university-trained engineers and scientists (Gooday, 1990; Schaffer, 1992; Falconer, 2014). Maxwell stated his strategy for mediating between Cambridge's gentlemanly culture and material experimental practice in his inaugural lecture:
> It will, I think, be a result worthy of our University ... if, by the free and full discussion of the relative value of different scientific procedures, we succeed in forming a school of scientific criticism, and in assisting the development of the doctrine of method (1890, p. 250).

His remarks about null methods must be read in the context of this drive to develop the "doctrine of method", a Kantian idea that Maxwell had encountered via William Hamilton and Alexander Bain (Cat, 2001, p. 401). For Maxwell, "method" connoted, "a non-arbitrary or non-personal procedure involving guidelines and desired outcomes… [that]… distinguishes it from obscure, untutored and unfounded opinion" (Cat, 2001, p. 430). The "doctrine of method" was the belief that such guidelines and procedures would form the proper basis for scientific criticism and the development of knowledge.

Maxwell now needed to bring experiment and measurement into his framework of guidelines and procedures. He had already adopted Bain's idea that similarity – or reasoning by analogy - was one of the bases for theoretical reasoning (Cat, 2001). To accommodate experiment, he now appears to have adopted two more of Bain's four fundamentals: classification and abstraction (Bain, 1856).[14] As Darrigol has observed, "Maxwell payed much attention to the nature of the physico-mathematical quantities, to their definition, and to their classification" (2003, p. 542). Both in the *Treatise,* and in the design of the Cavendish Laboratory itself, Maxwell categorised generic types of experiments and abstracted and classed the underlying principles (Maxwell, 1873a; Falconer, 2014). Maxwell's discussion of null methods, and the distinction he drew between registration or detection and measurement, formed part of this wider enterprise.

Although Maxwell and Thomson described Cavendish's as a "null method", Cavendish had not used this term, or any like it, himself. In the century following, a number of other electrical methods developed that were *later* identified as "null", most notably Ampère's experiments on the attraction between current-carrying wires (1821), Poggendorff's "compensation" method (1841), and Wheatstone's bridge (1843). As late as 1868, Latimer Clark described a number of null methods in his *Elementary Treatise on Electrical Measurement,* but did not use the term, or any synonym, and made no attempt to discuss them as a generic class. Instead, the book focused on particular instruments, and particular desired results, such as the specific conductivity of copper, or the location of faults.

An analysis of Google Books using Google's Ngram viewer indicates low or non-existent interest in null methods as a class until 1865-70 and a rapid rise thereafter (Figure 5). This chronology is in general agreement with Rutenberg's (1939) history of potentiometer (balance) methods.

---

[14] Bain's remaining fundamental was concept formation.

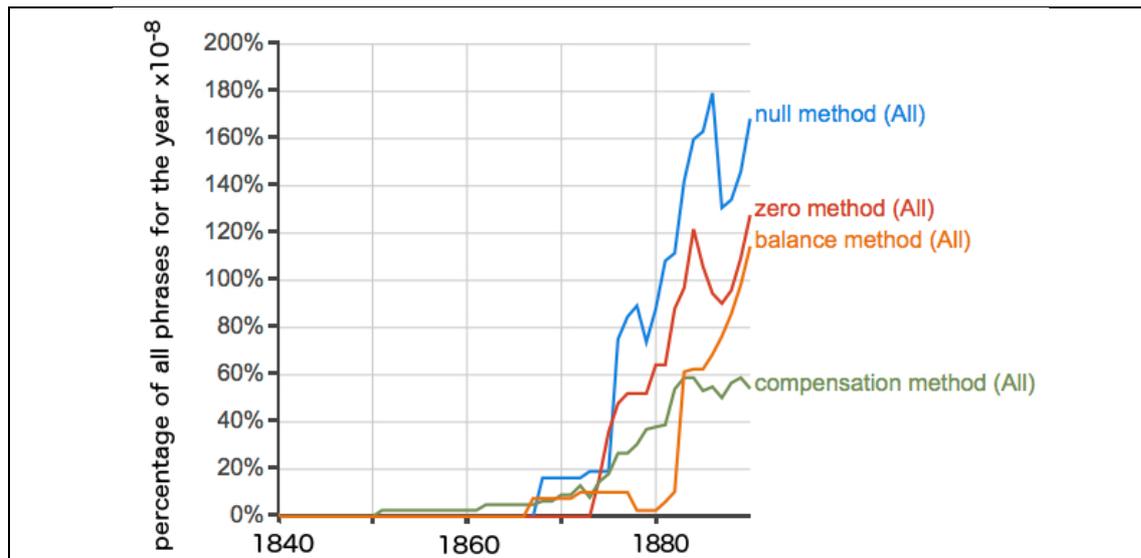

*Figure 5. Google Ngram for "null method" and synonyms from 1840 until 1890. Prior to 1840, Ngrams shows 0% usage of these phrases.*[15]

The term "null method" appears to have been popularised by Robert Sabine in his 1867 book *The Electric Telegraph*. Sabine described a number of these methods, for example Charles Victor de Sauty's use of a Wheatstone bridge for comparing the capacities of Leyden jars, "one of the most elegant of the many applications of the null-methods" (p. 379; de Sauty, 1865). Follow-up of the works picked out by the Ngram viewer suggests they come almost entirely from the field of telegraphy and electrical engineering. There is a strong positive correlation between the rise of the term "null method" and the growth of these fields.

Thus, when Maxwell discussed "null methods" in his *Treatise* in 1873, he was promoting a new, classificatory, way of thinking about experimental method, but one that was already beginning to rise among telegraph engineers. The methods themselves were not necessarily new, but treatment of them as a generic class, signified by a specific term, was. Maxwell had previously mediated between the practices of telegraph engineers and the concerns of mathematical physicists, via his 1863 paper with Fleeming Jenkin on ''The Elementary Relations Between Electrical Measurements'' (Hunt, 2015, p. 321; Mitchell, forthcoming). Now, in 1873, he adopted a similar strategy by assimilating telegraph engineers' null methods as a class to his doctrine of method.

Maxwell's treatment of null methods was inextricably linked with his assignment of electrical instruments into four classes: indicating instruments that were only "capable of detecting the existence of the phenomenon"; those intended to "register phenomena, not to measure them" because they gave readings that were consistent but bore no direct relation to the value of the quantity; a "higher class" whose scale readings were "proportional to the quantity to be measured so that all that is required for the complete measurement of the quantity is a knowledge of the coefficient by which the scale readings must be multiplied to obtain the true value of the quantity", and; "Absolute Instruments" which "contain within themselves the means of independently determining the true values of quantities" (1873a, p. 263).[16] A further principle of classification, "self acting" cut across these categories (e.g. 1873a, pp. 272, 274).

---

[15] The Google Ngram viewer plots the yearly count of a phrase, normalised by the number of phrases published in the year, in the corpus of books digitised by Google; Michel (2011). The output is best considered as indicating how widely a phrase was distributed among readers, rather than the number of authors using it, since books owned by several libraries result in multiple counts of the same text, as do the many journals that summarised longer papers. Faulty optical character recognition can give rise to anomalous counts at the level of detail, but counts for these particular phrases are sufficiently low that checking by hand was possible, and they appear reliable at the level of relative trends. For this reason a 5-year smoothing has been used in the plots.

Null methods provided a means of obtaining results using only the most basic class – indicating or detecting instruments. "Methods … in which the thing to be observed is the non-existence of some phenomenon, are called *null* or *zero* methods. They require only an instrument capable of detecting the existence of the phenomenon" (1873a, p. 263). Implicit was the argument that null methods had pragmatic advantages in an era when precisely calibrated instruments were expensive, and most electrical instrument manufacturers made little attempt at accuracy (Swinburne, 1888, p. 77). Null methods relieved the instruments – and their makers - of some of the burden of accuracy; instruments still needed to be sensitive, but need be neither highly consistent nor calibrated.

In making this instrumental distinction between detection and measurement Maxwell may have anticipated that null methods might localise issues that were developing around the concept of a physical quantity and its mathematical representation. Mitchell (forthcoming) discusses Maxwell's work on the representation of quantities and units. The quantitative claim of null experiments was of an absence of quantity. Null methods might thus provide universally valid results while avoiding debates around measurement units and standards – debates that Maxwell was, himself, heavily involved in as a member of the British Association Committee on Electrical Standards and as Cavendish Professor (Schaffer, 1992; Hunt, 1994, pp. 55-61; Gooday, 2004, esp. pp. 42-50; Mitchell, in press). We can only speculate where Maxwell might have taken such ideas, as his death precluded further development. But it seems likely that exploration of the measurement implications of null methods might form part of his developing framework of experimental principles and guidelines, furthering the doctrine of method.

As noted earlier, to put numerical limits on the precision of a null reading actually entailed attributing numbers to the sensitivity of the detecting instrument. Even then, Maxwell suggested a method that would give sensitivity as a ratio, avoiding the need to assign units to it. He described how to assess sensitivity by estimating the smallest deflection that could escape observation and comparing it with the deflection of an error of one percent (1873a, p. 396). This method was evident in his inverse square law experiment where he worked in ratios and mentioned no unit. In this special case, the experimental result was a dimensionless constant, the power of the distance.

The galvanometer sensitivity stressed by Maxwell was not the only relevant assessment. As he remarked, "… if there is no observable deflexion, then we know that the quantity … cannot differ from zero by more than a certain small quantity, depending on the power of the battery, the suitableness of the arrangement, the delicacy of the galvanometer, and the accuracy of the observer" (1873a, p. 395). He explicitly distributed the burden of accuracy between the design of the experiment, the apparatus and instruments, and the experimenter, and implicitly also over those who had designed and constructed the instruments. With null methods more generally, as with the inverse square law experiment, prior mathematical analysis was paramount in ensuring the "suitableness of the arrangement" and hence precision. For example, to compare resistances using a differential galvanometer, he went through a page of calculations to show the resistance of the galvanometer coils should be chosen to be one third of the resistance to be measured (1873a, p. 397).

---

[16] By "true value" Maxwell seems to have meant "absolute measure" in the sense that he, Thomson and Jenkin used it in the Second Report of the British Association Committee on Electrical Standards; Jenkin et al. (1873), p. 41; Mitchell (forthcoming), §2. To obtain the true value or absolute measure, some relation back to mechanical force or quantities of mass, length and time was required.

In a further classificatory move, Maxwell grouped null methods themselves into those arranged to produce a zero by the "opposing action of two currents", as in the differential galvanometer, and those where the zero was due to the "non existence of a current in the wire" (1873a, p. 396). He added a third method which, though "… not a null method, in the sense of there being no current in the galvanometer, it is so in the sense of the fact observed being the negative one, that the deflexion of the galvanometer is not changed when a certain contact is made" (1873a, p. 411); here registration rather than measurement was required. And here Maxwell stressed another known rationale for using null methods – by suitable bridge arrangements, variations in battery strength that might affect direct measurements had no effect on the galvanometer reading.

Within the context of the development of a doctrine of method, the null method for the inverse square law stands out as an oddity, since Maxwell's classification did not actually accommodate it. Although it might be considered an example of Maxwell's second type of method in which there was no force to be detected, no adjustment was required to achieve this result. *Pace* Thomson, there was no "seeking for a zero". Rather, the crux of the experiment was to demonstrate convincingly that the zero already existed naturally.

Yet within the wider rhetorical context at the time, claiming that the inverse square law was proved by a null method ensured relatively uncritical acceptance of both the law, and of the assertion of accuracy. It validated the mathematical-deductive approach, especially, perhaps, to telegraph engineers who already bought into the intrinsic reliability of null methods. Although the adverse impact of poorly adjusted resistances, inappropriate circuit arrangements, and insensitive galvanometers was frequently acknowledged (e.g. de Sauty, 1865; Heaviside, 1872), the rhetoric was almost entirely positive, centred on the method per se. This came first, with caveats entered only afterwards. Statements such as Oliver Lodge's on modifying Mance's method, "By this change it is converted into a strictly null method," or Schwendler's on Thomson's method of comparing capacities, "it is clear that this method must give exceedingly accurate results. For, in the first place, it is a perfectly null method …" took for granted a null method was the best and did not feel any need to justify their claims (Lodge, 1876, p. 145; Schwendler, 1878, p. 168). This remained the case, even when difficulties were obvious: "Prof. Hughes's second method of measuring [intermittent currents using a telephone in an induction balance]… is a much better one, because it is a null method and gives true readings, *though they are not easily interpretable*" (Lodge, 1880, p. 211, my emphasis). Thus, the wider context explored in this section provided a clear opportunity for Maxwell and the electrical programme through the doctrine of method.

7. **Conclusions**

Provided one accepts the inverse square law, Maxwell was undoubtedly correct in his assessment that Cavendish's null method was a more accurate demonstration than Coulomb's. However, in view of his previously expressed confidence in earlier null results, what was the value of repeating it?

Despite Maxwell's apparent confidence, analysis of contemporary textbooks shows that until around 1870, the inverse square law had been open to question, and its experimental base was far from robust. Following the death of Snow Harris in 1867, opposition to Coulomb's law decreased significantly in Britain. In the 1870s, Maxwell, Thomson, and Everett moved rapidly to close the door on alternative theories by constructing an experimental tradition of null tests that would be less open to critical scrutiny than Coulomb's experiments.

The power of the mathematical methods based on the law, and its role in securing their electrical programme, had raised the stakes for Maxwell and Thomson in protecting it. It gave them the confidence to assert the validity of the null test tradition with a cavalier attitude to earlier experimental arrangements and little critical inspection of its underpinning mathematical logic. His repetition of Cavendish's experiment represents the culmination of this effort for Maxwell, who died a few months later. He positioned his repetition at the head of both a mathematical tradition, from Pratt or Laplace, and the experimental one that he, Thomson and Everett, had constructed. The experimental tradition was, in turn, situated within his construction of a class of observational methods (null methods) that formed a step towards his developing "doctrine of method". This situation was an uneasy one philosophically, for it did not map neatly to his classificatory scheme, but it was a successful one rhetorically as it played into a widespread belief in the value of null methods – a belief for which Maxwell's promotion was partly, but not entirely, responsible.

A part of Maxwell's purpose may have been promotion of null methods as a class. He was pursuing the twin aims of developing the doctrine of method, and promoting the electrical programme in which he and Thomson were heavily engaged (Wise, 1995, pp. 135-136). The experiment instantiated many aspects of their programme, prioritizing mathematical deduction and careful design, and focusing attention on a precision instrument. However, precision was, to some extent, illusory. There was, according to Maxwell's own classification, no need for him to use the sophisticated Thomson's quadrant electrometer in the null experiment; a simpler detecting or registering instrument would have done, provided it was sensitive. But doing so protected Coulomb's law by ostensibly placing the burden of accuracy, via the perceived mathematical logic of the experimental model, on an instrument well able to bear it, while deflecting attention from less trustworthy aspects of the method. There was no need to push the experiment to the limits of precision, partly because of the prior success of aspects of its positioning, and partly because Maxwell viewed it as an exemplar of the doctrine of method as much as an affirmation of the inverse square law.

Null methods were, generally, fairly uncritically conceived as offering precision. As Maxwell stressed several times, they required only "detection" rather than "measurement". This was not an idle distinction for someone whose "insights into measurability were deep" even if only sporadic (Darrigol, 2003, p. 545), and who took care near the beginning of the *Treatise* to establish that electrification was a properly measurable quantity (Maxwell, 1873a, p. 35). Maxwell's use of this distinction, and its interaction with his views on the relations between physical laws and the laws of numbers, which authors such as Boumans (2005) have seen as a forerunner of the representational theory of measurement, deserve further exploration.

In Thomson and Maxwell's hands, promotion of null methods increased the subordination of experimental to mathematical physics in Britain. However, we should be cautious of accepting the generality of this conclusion too easily. The interaction of method and promotion with context was all-important. Mathematical physics was increasingly dominant in Britain, Germany, and France. But in Germany, null methods are not evident in Cahan's (1990) study of Kundt's efforts to prioritise theory. And in France, Mitchell (2012) shows how Lippmann mobilised null methods *against* mathematical physics, attempting to shed measurement of theory.

Although the null method may have avoided issues about standards, the inverse square law itself was deeply implicated in the drive to establish such standards. That Maxwell did develop an experiment he avowed already so well established, emphasises once again its importance to the whole electrical enterprise. His statement, "no actual measurement of force was required" (Harman, 2002, p. 539), may have been pure rhetoric, but through a series of conceptual and historical re-orderings, he succeeded in establishing the law and a method upon which its demonstration has been based ever since.

### Acknowledgements

I have enjoyed many stimulating discussions with Daniel Mitchell while writing this paper, and have also benefited greatly from the valuable and insightful comments of two anonymous referees. I am grateful to the Radio Astronomy Group at the Cavendish Laboratory, Cambridge, for hospitality while conducting the research.